# Towards sustainable transport: wireless detection of passenger trips on public transport buses

Vassilis Kostakos
Lab:USE, University of Madeira
Human Computer Interaction Institute, Carnegie Mellon University
vassilis@cmu.edu

## Abstract

An important problem in creating efficient public transport systems is obtaining data about the set of trips that passengers make, usually referred to as an Origin/Destination (OD) matrix. Obtaining this data is problematic and expensive in general, especially in the case of buses because on-board ticketing systems do not record where and when passengers get off a bus. In this paper we describe a novel and inexpensive system that uses off-the-shelf Bluetooth hardware to wirelessly detect and record passenger journeys. Here we show how our system can be used to derive passenger OD matrices, and how our data can be used to further improve public transport services.

## Introduction

More than 20% of the world's energy is spent on transportation.[1] At a time when the environmental implications of modern life are scrutinised, reducing the energy spent on transport is a key objective for achieving sustainability for our way of life. With more than 50% of commuters driving their own car to work [1], governments are actively campaigning for the use of public transport. Interestingly, researchers point out that even if more passengers choose public transport, the reduction in energy consumption will not be considerable due to the inefficiencies of public transport:

*"Trains and buses are potentially much more efficient than cars, if only they were full. But the way we do public transport at present, trains and buses are not that much more energy-efficient than cars. There remain many other good reasons for encouraging a switch to public transport*

---

[1] International Energy Outlook 2007, United States Department of Energy - Washington, DC. http://www.eia.doe.gov/oiaf/ieo/index.html. Retrieved on June 2, 2008.

*(for example avoiding congestion and reducing accidents), but don't expect to reduce energy consumption enormously by a switch to public transport. "* [2, p. 133].

In Figure 1 we present a summary of the energy cost of various modes of transportation as described in [2]. While developing more efficient bus engines is an obvious way to improve buses' energy efficiency, an orthogonal approach is to consider ways of increasing bus occupancy. Designing a more efficient public transport network, where more seats are occupied more often, can greatly reduce the total energy spent for each passenger and hence bring us closer to achieving sustainable public transport.

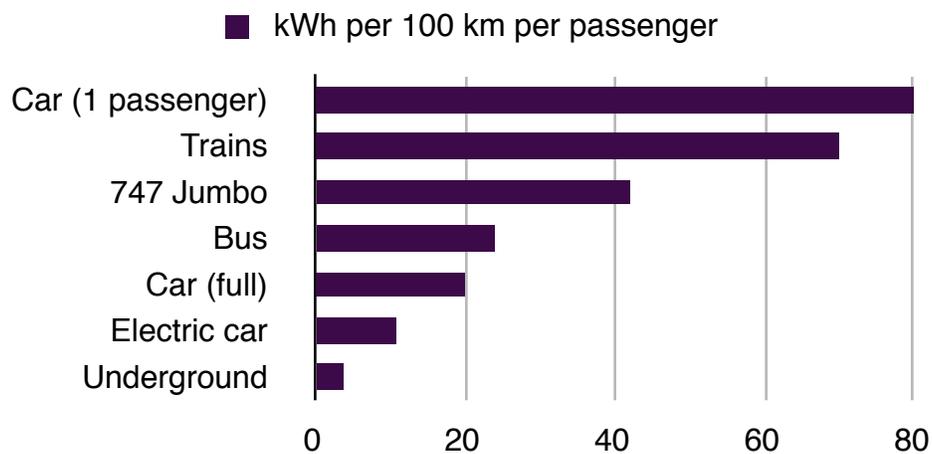

**Figure 1: Typical energy consumption of public transport. Notes: Car consuming 33mpg. Trains travelling at 33km/h, and considering energy cost for lighting, lifts, depots and workshops. 747 Jumbo at cruise speed of 900km/h. Underground average speed of 48km/h. [2]. Bus carrying 14.4 passengers at average speed of 18 km/h, distance between stops 0.3 km [3].**

A key requirement for designing and improving the public transport network's efficiency is obtaining an Origin/Destination (OD) matrix for passenger movement. Such a matrix is effectively a table that describes the flow of passengers between various points in the transport network (or alternatively between points on a map). In practice, the cumulative OD matrix can be filtered in many ways, e.g. to display the flow of passengers during peak hours or weekends. The design of public transport networks draws heavily on this information, and many decisions such as the scheduling of buses and drivers [4] are directly based on this information.

Traditionally, the process of obtaining an OD matrix has been laborious and expensive because it typically involves human observers manually counting the number of passengers over a number of days. More recently, the use of electronic ticketing systems has greatly simplified this process. While some public transport systems such as the underground have an end-to-end ticketing system (i.e. the system records passengers' entry and exit points), others do not. For instance, most ticketing systems on buses record when passengers get on a bus, but do not record when they get off the bus. As a result, bus services still employ manual observation to capture the OD matrix, or rely on expensive sensor systems that can count the number of passengers on a bus at any time without being able to detect the specific origin and destination of any single passenger.

A further complicating factor is that depending on the amount of growth and change within a city the OD matrix changes over time, while in extreme cases (such as the Olympic games) these changes can happen abruptly [5]. Hence, while human observation can be used to capture an OD matrix, this information may become inaccurate or obsolete within a few months. Depending on the frequency of human observations (typically every 6 months due to high cost), the bus service is expected to operate with a certain degree of inefficiency measured in terms of occupancy (percentage of seats occupied at any given time). An up-to-date OD matrix can help public transport authorities to better allocate their resources (drives, busses, repair crews), develop a more efficient transport network, and to fine-tune the operation of their network, effectively reducing the total energy spent per passenger. Furthermore, authorities can use such information to refine their reward schemes.

**Outline**

In this paper we focus on the development of a system that automates the process of capturing the passenger OD matrix along a specific line or zone. Here we describe a novel and inexpensive system that uses off-the-shelf Bluetooth hardware and passengers' Bluetooth-enabled devices to accurately record passenger journeys, with no need for special software to run on passengers' devices. Our system uses constant Bluetooth discovery as a mechanism to record when a passenger boards and exits a bus. Combining this information with the on-board localisation system, we are able to determine the exact bus stop where a passenger boards and exists a bus.

# Related work

Most research considering OD matrices focuses on deriving accurate estimations from incomplete data (for an overview, see [6]). Typically, human observations and traffic counts do not cover every single segment of the transport network, and hence estimates of the flow of passengers for unobserved parts of the network are statistically derived [7]. Furthermore, some observation schemes do not rely on direct observation but rather on passenger questionnaires, hence introducing unreliability in the data.

In cases where an automated ticket system exists, OD matrices can be captured from ticketing data. It is important, however, to note that not all automated ticket systems are suitable for this tasks. For instance, in most cases buses do not record passengers' exit points, hence capturing incomplete information about journeys. Furthermore, many ticket systems were not originally designed for data collection [8]. As a result, many lack important information, collect data for specific and limited purposes, and record in a fragmented, intermittent, and difficult-to-use format. Furthermore, different subsystems (e.g. GIS and ticketing systems) may be supplied by different vendors and thus managed in completely different databases, thus making analysis difficult.

The problem of inferring an OD matrix from origin-only data has been addressed by Zhao et al. [8] in their analysis of the Chicago Transit Authority rail system which collects origin-only data. However, their analysis is based on a number of assumptions (p. 381):

1. There is no private transportation mode trip segment (car, motorcycle, bicycle, etc.) between consecutive transit trip segments in a daily sequence;

2. Passengers will not walk a long distance to board at a different rail station from the one where they previously alighted.

3. Passengers end their last trip of the day where they began their first trip of the day.

Such assumptions inevitably introduce inaccuracies in the calculated OD matrix, especially when considering a bus-only network. Furthermore, this approach completely fails to take into account one-way tickets, and passengers who do not have a permanent travel card.

Mobile phone tracking has been used as an approach to measure the flows of passengers between parts of a city [9]. Such data, however, has low spatial resolution and is most appropriate for long-distance segments such as highways. This approach cannot be effectively used in a condensed network such as inner-city bus networks.

Another approach to capturing passenger trips on busses is to make use of the onboard cameras and apply automated head detection [10]. While this approach can be used to track passengers getting on/off the bus with some degree of accuracy, this system does not differentiate between passengers and is expensive in terms of equipment and computation. Some commercial solutions rely on pressure sensitive carpets or infrared sensing for capturing the number of passengers onboard a bus at any given time.[2]

Finally, while our use of Bluetooth as a means of capturing OD matrices is novel, Bluetooth has been used on public busses before. Most typically, Bluetooth is used for automatic downloading of diagnostics and reports once busses return to their garage.[3] Furthermore, Bluetooth has been considered as a replacements for cables, which can run up to 4 km on a single bus [11], thus reducing weight and overall petrol consumption. In addition, prototype systems have considered Bluetooth as a mechanism for providing passengers access to the internet [12]. Finally, other systems have considered exploiting passengers' mobile devices for optimising the transport network, by exploring how passengers' mobile devices can help plan and execute journeys in realtime [13]. This approach, however, requires custom software to run on passengers mobile devices, which introduces considerable development costs and compatibility issues.

---

[2] http://www.acorel.com

The above uses of Bluetooth technology onboard buses are encouraging for us, because we can easily piggy-back our system on top of any existing Bluetooth infrastructure. We now proceed to describe our system.

## Implementation

Our system was implemented for Horários do Funchal, the public transport operator in Funchal, Madeira, Portugal. This organisation has over 160 buses serving about 30 million passengers per year, across more than 1400 bus stops. These buses have an elaborate localisation and ticketing system, which was in use prior to our study.

Each bus is equipped an on-board GPS system, complemented by a digital odometer (distance travelled) and door sensor (doors open/closed). These three components are used to determine the bus location at any given moment. Buses report their location using a GPRS connection, and all bus locations are fed centrally into a real-time commercial simulator that estimates when each bus will reach the next bus stop. These estimates are then transmitted to bus stops using GPRS, and each bus stop displays the estimated arrival time of each service on an electronic display.

Additionally, each bus has a ticketing system that records information about the time when passengers boarded the bus, and the type of ticket they purchased. Horários do Funchal uses RFID tickets for all passengers, including those purchasing single trips. This data is stored on-board and transmitted using WiFi each time the bus returns to the central garage.

Our system was developed on a Gumstix Waysmall btx-400, which has a 400MHz processor, a class 1 Bluetooth adapter, and 16 MB of storage. We refer to our system as a "scanner", and for our trials we installed one scanner in one bus. The scanner was installed on the roof inside the main bus cabin, near the exit area located in the centre of the bus (see Figure 2).

---

[3] Nohau Elektronik, Bluetooth in Public Transportation Bus Fleet of Milan. http://whitepapers.silicon.com/0,39024759,60113218p,00.htm

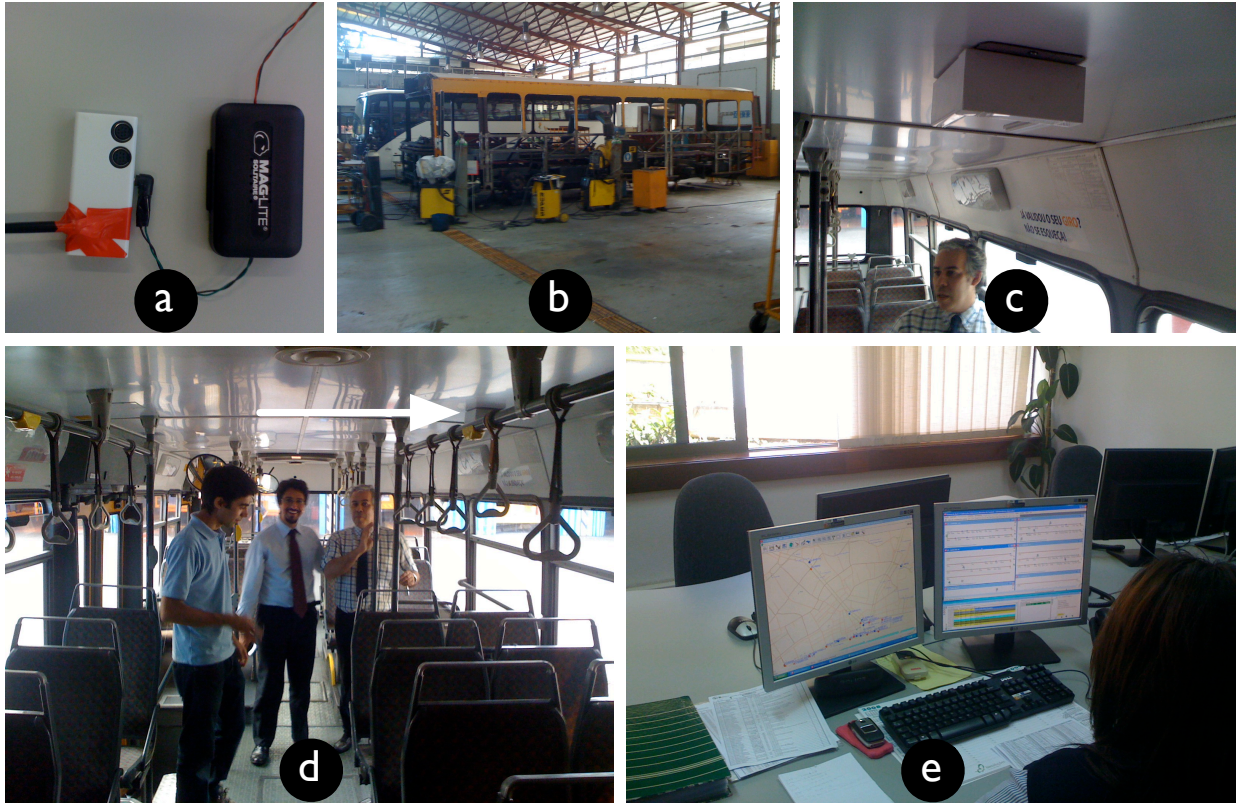

**Figure 2: Installation of our system. a: The Gumstix computer (left), along with at 24 to 5 volt converter (right) used to power the Gumstix with the bus' electric circuit. b: a bus being rewired. c: our final installation consisted of a protective plastic case attached to the roof of the cabin. d: the system (indicated with an arrow) is installed near the centre of the main cabin. e: the control centre where real-time data is gathered from the buses' localisation systems and the whole operation is overseen.**

The scanner software is rather basic: it constantly issues a Bluetooth discovery request and records the results. According to the standard Bluetooth protocol, a Bluetooth device set to "Discoverable" mode must respond to the discovery request by transmitting its unique Bluetooth identifier (12 hex digits) and device class (6 hex digits). Our scanner constantly issues the same discovery request, and constantly records the presence of the various devices it encounters (along with the date and time of each distinct instance a device was discovered). Using this approach, we have the additional benefit of not requiring any special software to run on passengers' devices. The only requirement is that passengers set their devices' Bluetooth adapter to "Discoverable" mode. While we had not explicitly measured the proportion of residents in Funchal carrying a discoverable Bluetooth device prior to our study, we expected this proportion to be in the order of 7.5% of the population, as described in [14].

## Data analysis

We deployed our scanner for four weeks. During this period the bus covered 19 different routes at different times of the day. This is due to the way buses, drivers and routes interweave in the schedule of Horários do Funchal in order to improve operational robustness. Practically, this means that our scanner collected data for a number of different routes while remaining on the same bus. During our trial the scanner recorded more than 1000 unique Bluetooth devices.

In Figure 2 we present how we correlate the two datasets we have access to: our Bluetooth data and the bus localisation data. First, we pre-process our Bluetooth data such that we derive device "trips". A device trip is defined by the unique Bluetooth ID of a device, the time when the device become visible to our scanner, and the time when the device disappeared from our scanner. In practice, our scanner discovers nearby devices every 3-10 seconds. To derive device trips we accumulate consecutive device discoveries that are less than 5 minutes apart. We set such a high threshold to compensate for instances where standing passengers possibly block the Bluetooth signals onboard the bus.

Having derived device trips, we then correlate these trip times with the bus localisation database. By analysing the localisation data, we were able to calculate the exact times when the bus visited the bus stops on its route (with a 10 second error margin). This event is recorded when the bus reaches the bus stop and the driver opens the doors. This way, we were able to identify the exact bus stop when a device first appeared (hence the passenger boarded the bus), and when a device disappeared (hence the passenger exited the bus).

We should note that the correlation process removes a lot of noise from our Bluetooth dataset. For instance, our scanner detected devices while the bus was out of service or being repaired. Without the localisation data, there is not way to verify if such Bluetooth data reflect passengers or not. With the localisation data, however, we know that these devices appeared when the bus was not en-route, hence we can discard this data. Similarly, if our scanner picked up devices from outside the bus (e.g. passengers waiting at a bus stop), then these devices would appear to board and exit at the same bus stop, hence can easily be identified and removed. Finally, if any non-passenger devices where picked up in-between bus stops then our correlation process would not assign them any bus stop at all. We also note that an assumption we make in our analysis is that passengers do not enable or disable their Bluetooth device while onboard a bus.

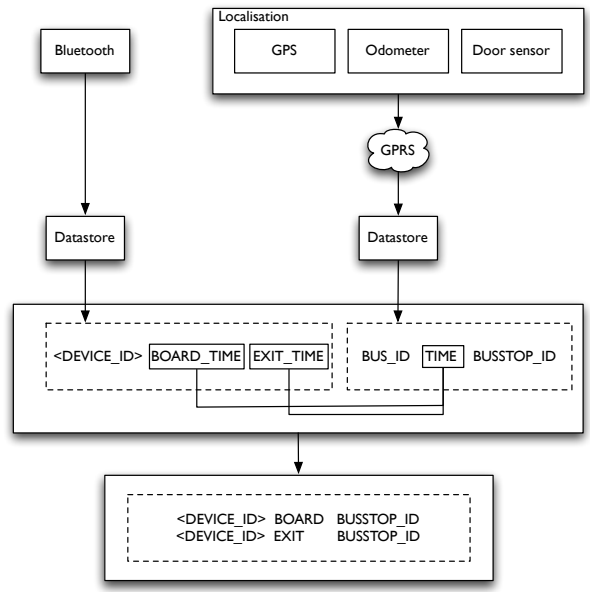

**Figure 3.** Correlating the Bluetooth data with the bus localisation data. From our Bluetooth data we calculate the times when a device boarded and exited the bus. Using the bus localisation data, we are able to figure out at what times the bus visited each of the bus stops on its route. Combining these two datasets, we are able to calculate the bus stops where a device boarded and exited the bus.

Next, we present a brief overview of the passenger behaviour as recorded by our scanner. Here we only focus on the novel analyses that are enabled by our system - in other words analyses that require knowledge of individual passengers' destination. More traditional types of analysis, such as calculating how often passengers use a service, can be carried out using origin-only data. In Figure 4 (top) we present the average number of passengers on board the bus throughout the day. This graph correctly highlights the expected morning, mid-day and evening peaks, representing people going to work, to lunch and returning home respectively.

Additionally, we examined the correlation between our Bluetooth data and the electronic ticket data (inset in top of Figure 4). We found a correlation of $R^2 = 0.737$ between the number of device trips per hour and the number of tickets validated per hour on our bus. Additionally, the slope of the correlation is closest to 1 when the number of device trips is multiplied by a factor of 10.26. This suggests that about 9.7% of passengers have Bluetooth-discoverable devices, which is very close to our original estimate of 7.5% derived from [14]. Interestingly, the increase in morning ticket activity is under-registered by our Bluetooth scanner, suggesting that the bus may be too congested for Bluetooth to function without interference.

We also show a breakdown of the duration of passenger's trips (bottom of Figure 4). It is interesting to note that many passengers make small trips. This is possibly due to the geography of Funchal which features many steep hills. As a result, passengers are motivated to take the bus even for very few stops because of the steep terrain.

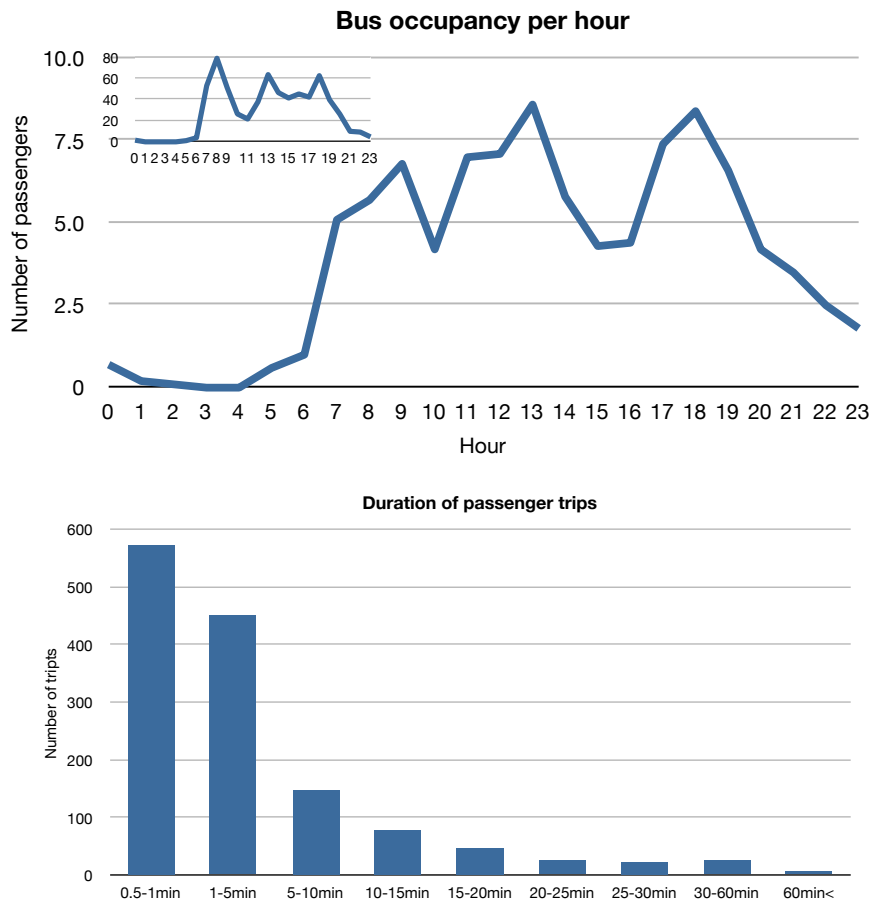

**Figure 4. Top: the average number of passengers onboard the bus at any hour during the day (inset: number of tickets validated per hour). Bottom: a histogram of passenger trip duration.**

Finally, we were able to derive the passenger OD matrix for the lines that our bus covered during our trial. In Table 1 we present only a segment of this matrix, since the complete matrix is too large to reproduce fully on paper. Here we show the passenger trips for a subset of the bus stops on route 138, and additionally considering outward services only. Again we see that most trips are made between consecutive bus stops, and we can easily identify that the most popular segment in this route is between bus stops S303 and S305.

| Origin | Destination | | | | | |
|---|---|---|---|---|---|---|
| | S297 | S299 | S301 | S303 | S305 | S307 |
| S297 | - | 2 | 2 | - | - | - |
| S299 | - | - | 3 | 3 | - | - |
| S301 | - | - | - | 1 | - | - |
| S303 | - | - | - | - | 8 | - |
| S305 | - | - | - | - | - | 3 |
| S307 | - | - | - | - | - | - |

**Table 1. An excerpt of the passenger OD matrix for service 138. The leftmost column lists the origin bus stops, while the subsequent columns represent the destination bus stops.**

## Discussion

### Benefits

A technical account of how an OD matrix can be used to redesign a public transport network is beyond the scope of this paper. Indicatively, however, we highlight some interesting points from such an analysis. An OD matrix is typically considered in terms of zones, and in the case of Horários do Funchal the OD matrix has about 20 zones. This data is then used in conjunction with traffic analysis and knowledge of the city to set the parameters of the transit network such as the routes between zones and the timing of services. Additionally, the OD matrix can be used to identify locations that attract many passengers and, combined with the relative distance between those locations, identify optimal ways of linking such locations. An array of tools can analyse an OD matrix and derive network improvements and optimisations, considering the micro, meso and macro scales of transportation (for an overview, see [15]). Finally, the OD matrix can be used to optimise network simulators used to predict when buses will reach the next bus stop, as well as optimise the schedule itself, by better allocating buses and drivers to routes [8].

In this sense our system provides rich data about origins and destinations, as we record data on a bus-stop level rather than zone level which can be larger than a square kilometer. This means that our OD matrix can be used to consider bus occupancy for every route in the network. Similar to Figure 4 (top), each route can be analysed for occupancy per hour, as well as per day of week. This can help identify under-utilised and over-utilised services, hence guiding decisions on merging or cancelling routes, increasing the frequency of existing routes, or introducing new routes.

While our OD data is richer, both in terms of higher geographic precision and more longitudinal datasets, how much improvement can we expect? According to the American Public Transport Association, up to 70% of passenger trips are related to work and school hence are highly predictable.[4] While this may be the case, a survey-based is extremely slow in responding to changes in cities. New roads, schools, sports facilities, underground stations, policies and the climate are some of the factors that have a continuous impact on passenger trips. Hence, while up to 70% of trips are highly predictable in the short-run, transport authorities need monitor the differences over time and react accordingly. At worst, our system can act as an early warning in changes to passenger behaviours and needs. At best, our system may help authorities uncover subtle patterns in passenger behaviour, and perhaps predict shifts in such behaviour.

**Limitations**

An important issue to discuss is the penetration of Bluetooth through our passenger community. Our analysis assumes that Bluetooth-carrying passengers are a random sample of the population, but it can be argued that people carrying Bluetooth are most likely well-off individuals and teenagers. Since our study has taken place in a real-world setting, it is impossible for us to know the demographics of those carrying Bluetooth and consequently know if our data is skewed in any way. What we do point out, however, is that mobile phones are increasingly commoditised and accessible to larger parts of the populations, with Bluetooth being a technology that is increasingly considered a standard feature of mobile phones. Hence, while at the moment we cannot accurately assess which passengers carry (and enable) Bluetooth, our assumption is that in time the portion of Bluetooth-carrying passengers will grow steadily.

In addition, the low cost of our system (1/20th the cost of commercial passenger-counting systems) makes it possible to install it on more buses. Hence, while our system only detects Bluetooth-carrying passengers, it can do so on more services. Considering the business perspective of Horários do Funchal, lets assume they can afford to install a commercial passenger-counting technology on 5 of their 160 busses, hence recording 5/160ths of their annual passengers. With the same cost, they can use our system to cover 100 busses that record about 10% of 100/160ths of their passengers, or 10/160ths of their annual passengers, which is double the passengers compared to commercial systems. What is likely to be a winning strategy, however, is to use a mixture of the two systems, hence obtaining both fine-grained and high-volume data.

---

[4] http://www.apta.com

**Privacy**

Our use of Bluetooth has privacy implication which are increasingly becoming aparent to users. Our system lets us track individual passengers' behaviour over time, and by consequence records very precise information about people's location at any give time. Such information, should it fall in the wrong hands, can be used for harmful intents and purposes. For instance, a culprit may use knowledge of the fact that Alice is currently on the bus to infer that she is not at her flat and rob it. Hence, bus companies need to make sure that Bluetooth data is stored securely and is inaccessible to third parties. We should also point out that this strategy ought to be followed for magnetic and RFID tickets as well, since they can also be used in the exact same way to infer the location of passengers over time. In this case, a feature of Bluetooth that works to its advantage is that the Bluetooth ID cannot be linked to users' identity, unlike magnetic and RFID long-term passes which are usually linked to people's identity when they are issued. Furthermore, users can always disable their Bluetooth, thus avoiding detection.

**New uses**

In addition to the above approaches to utilising our data, we are also exploring new types of analyses enabled by our system. These are primarily aimed at improving the bus service from the users' perspective by providing them with a better and customised service. Our approach is to consider individualised OD matrices, i.e. to generate an OD matrix for each unique Bluetooth device. During our trials we recorded more than 1000 distinct devices. For each device we are able to calculate a customised OD matrix, which helps us predict where each passenger is likely to want to go and when they will get there, given the bus stop at which they are standing, the day of week, and time of day. Additionally, our system lets us identify groups of passengers that travel together, and groups of passengers that encounter each other on the bus on a regular basis. Such data is useful for getting deeper understanding of people's transport habits, and furthermore is essential for defining appropriate commercial strategies which include the creation of specific fares (e.g. friend tickets) .

Considering individualised passenger information is an approach that can enable new types of services for before, during and after a passenger's trip. Bluetooth-augmented bus stops can use individualised OD data to identify where waiting passengers are likely to want to go. Thus, bus stops can display the time when the next bus will arrive as well as when the bus will reach the bus stops of interest to the waiting passengers. Knowledge of where the waiting passengers want to go can also be used to deliver relevant information about events and attractions. While onboard a bus, our Bluetooth infrastructure can be used to calculate relationships between passengers, such as how much time passengers spend together or how often they encounter each other. Such information can potentially be used to create applications and games to keep passengers engaged during their trip. For passengers who ultimately reach their destination, a nearby bus stop can once again predict where and when the passengers may want to go next, and display relevant information.

## Conclusion and ongoing work

The motivation for our work has been to improve public transport's energy efficiency. As we describe in the introduction, a key factor contributing to public transport's current inefficiency is low seat occupation. This problem can be addressed by having accurate and frequently-updated OD matrix data to support an efficient and timely service, something which is currently too expensive for bus companies. In turn, we chose to improve this data collection process by developing a system that cheaply and automatically collects more data about passengers' travel behaviour than previously possible. The major output of our work has been the passenger OD matrix, and the subsequent graphs and analyses that can be derived from it.

Fulfilling our goal of improving the transport network, and ultimately improve its energy efficiency, is a long process involving multiple stake-holders. However, the transport engineers at Horários do Funchal have been very positive about our results, and we are in the process of extending our system to more buses. As part of this expansion, we are considering ways of linking our system to the on-board localisation system and using its GPRS connection to remotely collect our data in real time. As Bluetooth technology is increasingly being used onboard busses we hope that our system can be used by more transport organisations to collect data about their passengers and eventually improve their operation.

## Acknowledgements

The author wishes to thank Claudio Materno, Filipe Santos, Alfredo Pereira, Jonathan Vieira, Jeronimo Faria and Mayuree Srikulwong. This work is supported by Horários do Funchal, and FCT Portugal via the CMU-Portugal agreement.